\documentclass[%
 reprint,
showpacs,
 amsmath,amssymb,
 aps,
 prl,
]{revtex4-1}

\usepackage{graphicx}
\usepackage{dcolumn}
\usepackage{bm}
\usepackage{hyperref}
\usepackage{units}
\usepackage{feynmp}
\usepackage[caption=false]{subfig}
\usepackage{tensor}

\DeclareMathOperator{\tr}{tr}
\newcounter{myequation}
\makeatletter
\@addtoreset{equation}{myequation}
\makeatother

\begin{document}

\preprint{APS/123-QED}

\title{Non-analytic behavior of the Casimir force across a Lifshitz transition in a spin-orbit coupled material}

\author{Andrew A. Allocca}
 \affiliation{Joint Quantum Institute and Condensed Matter Theory Center, Department of Physics, University of Maryland, College Park,
  Maryland 20742-4111, USA} 
  
\author{Justin H. Wilson}
 \affiliation{Joint Quantum Institute and Condensed Matter Theory Center, Department of Physics, University of Maryland, College Park,
  Maryland 20742-4111, USA}

\author{Victor Galitski}
 \affiliation{Joint Quantum Institute and Condensed Matter Theory Center, Department of Physics, University of Maryland, College Park,
  Maryland 20742-4111, USA}

\date{\today}

\begin{abstract}
We propose the Casimir effect as a general method to observe Lifshitz transitions in electron systems. 
The concept is demonstrated with a planar spin-orbit coupled semiconductor in a magnetic field. 
We calculate the Casimir force between two such semiconductors and between the semiconductor and a metal as a function of the Zeeman splitting in the semiconductor. 
The Zeeman field causes a Fermi pocket in the semiconductor to form or collapse by tuning the system through a topological Lifshitz transition. 
We find that the Casimir force experiences a kink at the transition point and noticeably different behaviors on either side of the transition. 
The simplest experimental realization of the proposed effect would involve a metal-coated sphere suspended from a micro-cantilever above a thin layer of InSb (or another semiconductor with large $g$-factor). 
Numerical estimates are provided and indicate that the effect is well within experimental reach.
\end{abstract}

\pacs{12.20.-m, 71.70.Ej, 73.61.Ey, 75.70.Tj}
\maketitle

\begin{figure}[b]
\centering
\includegraphics[width=0.6\columnwidth]{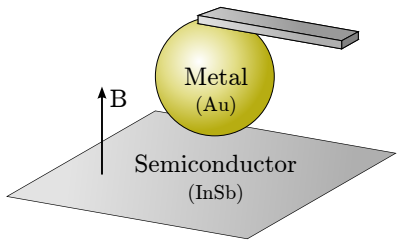}
\caption{\label{fig:CasimirExperiment} The geometry typically used in experimental measurements of the Casimir force is a gold coated sphere suspended above a planar plate from a cantilever. We consider a lower plate of indium antimonide with an applied magnetic field.}
\end{figure}

In 1948, Casimir predicted attraction between two neutral, perfectly conducting materials \cite{Casimir:1948dh}, and after nearly fifty years of theory \cite{Milton:9_Dy4-Ik}
, experimental evidence was presented by Lamoreaux \cite{Lamoreaux:1997ui}. 
Following this discovery there was a flurry of theory \cite{Bordag:2001tt,*Milton:2008kh} and experiment  \cite{Lamoreaux:2005kd,*Mohideen:1998wf,*Munday:2009vk,*Banishev:2012df} which led to an astounding amount of theoretical and experimental machinery.
With this machinery, one can use the Casimir force as a probe of real material properties -- e.g., correlations, phase transitions.
In this letter, we consider how the Casimir force changes across a Lifshitz transition. 
The model describes a thin layer of indium antimonide and could be experimentally realized in the common experimental setup for Casimir measurements as shown in Fig.~\ref{fig:CasimirExperiment}.

The Casimir effect for real materials, as first developed by Lifshitz \cite{Lifshitz:1956vv,*Dzyaloshinskii:1961ew}, explicitly depends on the electromagnetic response of a material. 
This response can be built into the boundary conditions of the electromagnetic field itself. 
\begin{fmffile}{Casimir-diagrams}
Diagrammatically, the Casimir energy between two plates $A$ and $B$  (from which the force is derived) takes the schematic form
\begin{align} \label{eq:schematic-casimir}
\mathcal{E}_c & = 
\parbox{17mm}{
\begin{fmfgraph*}(60,55)
\fmftop{t}
\fmfbottom{b}
\fmfleftn{i}{2}
\fmfrightn{o}{2}
\fmf{phantom,tension=3}{t,d1}
\fmf{phantom,tension=1}{i2,l1}
\fmf{phantom,tension=1}{r1,o2}
\fmfpoly{empty,smooth,label=$A$,tension=1.5}{l1,d2,r1,d1}
\fmf{phantom,tension=2}{d2,d3}
\fmfpoly{empty,smooth,label=$B$,tension=1.5}{l2,d4,r2,d3}
\fmf{phantom,tension=1}{i1,l2}
\fmf{phantom,tension=1}{r2,o1}
\fmf{phantom,tension=3}{d4,b}
\fmffreeze
\fmf{photon,right=0.3}{l1,l2}
\fmf{photon,left=0.3}{r1,r2}
\end{fmfgraph*}
}
+
\parbox{28mm}{\centering
\begin{fmfgraph*}(60,60)
\fmftopn{t}{2}
\fmfbottomn{b}{2}
\fmfleftn{i}{2}
\fmfrightn{o}{2}
\fmf{phantom,tension=5}{t1,d1}
\fmf{phantom,tension=1}{i2,l1}
\fmf{phantom,tension=1}{r1,l3}
\fmfpoly{empty,smooth,label=$A$}{l1,d2,r1,d1}
\fmf{phantom}{d2,d3}
\fmfpoly{empty,smooth,label=$B$}{l2,d4,r2,d3}
\fmf{phantom,tension=1}{i1,l2}
\fmf{phantom,tension=1}{r2,l4}
\fmf{phantom,tension=5}{d4,b1}
%
\fmf{phantom,tension=5}{t2,d5}
\fmf{phantom,tension=1}{r3,o2}
\fmfpoly{empty,smooth,label=$A$}{l3,d6,r3,d5}
\fmf{phantom}{d6,d7}
\fmfpoly{empty,smooth,label=$B$}{l4,d8,r4,d7}
\fmf{phantom,tension=1}{r4,o1}
\fmf{phantom,tension=5}{d8,b2}
\fmffreeze
\fmf{photon,right=0.3}{l1,l2}
\fmf{photon,left=0.3}{r3,r4}
\fmf{photon}{r1,l4}
\fmf{photon}{r2,l3}
\end{fmfgraph*}
}
+ \cdots  
\end{align}
where 
$\parbox{7mm}{\centering
\begin{fmfgraph*}(25,10)
\fmfleft{i}
\fmfright{o}
\fmf{phantom,tension=6}{i,v1}
\fmfpoly{empty,smooth,label=$X$}{v1,v2}
\fmf{phantom,tension=6}{v2,o}
\end{fmfgraph*}
} = \parbox{10mm}{\centering
\begin{fmfgraph*}(25,10)
\fmfleft{i}
\fmfright{o}
\fmf{phantom,tension=6}{i,v1}
\fmfpoly{empty,smooth}{v1,v2}
\fmf{phantom,tension=6}{v2,o}
\end{fmfgraph*}
} (1 + \parbox{13mm}{\centering
\begin{fmfgraph*}(40,10)
\fmfleft{i}
\fmfright{o}
\fmf{phantom,tension=6}{i,v1}
\fmf{photon,tension=2}{v1,v2}
\fmfpoly{empty,smooth,label=$X$}{v2,v3}
\fmf{phantom,tension=6}{v3,o}
\end{fmfgraph*}}
)$ is the dressed current-current correlation function for plate $X$ while \!\parbox{7mm}{\centering
\begin{fmfgraph*}(25,10)
\fmfleft{i}
\fmfright{o}
\fmf{phantom,tension=6}{i,v1}
\fmfpoly{empty,smooth,label=$X$}{v1,v2}
\fmf{phantom,tension=6}{v2,o}
\end{fmfgraph*}
}
is the usual current-current correlator derived in linear response theory -- a material dependent quantity related to conductivity. 
It enters the expression in a crucial way, and thus, features in the frequency-dependent conductivity translate to features in the Casimir force. 
\end{fmffile}
Being able to tune the Casimir force by modifying a material's electromagnetic response would have important applications for precision gravity experiments \cite{Bordag:1998jg,*Long:2003ca,*Decca:2005hf,*Kapner:2007hh,*Weld:2008ei} and applications to nanotechnology \cite{Capasso:2007fl}. 

From the other direction, and importantly for the subject of this paper, any change of the Casimir force would be an indication of a change in the material's properties. 
Special geometries \cite{Levin:2010ft} and boundary conditions \cite{Fulling:2007hh} can change the Casimir force to be repulsive, though with symmetric geometries without time-reversal symmetry breaking, one can not escape an attractive effect \cite{Kenneth:2006fm}. 
Just as a repulsive effect would be a signature of some time-reversal symmetry breaking (such as in the case of two quantum Hall plates \cite{Tse:2012hv} or topological insulators with gapped surface states \cite{Grushin:2011bz}), other changes in the Casimir force can be attributed to other material properties.
For instance, Bimonte and coauthors showed that one can in principle measure the change in Casimir energy between a normal and superconducting state \cite{Bimonte:2005ek,*Bimonte:2005fh}. 
Additionally, the critical Casimir effect \cite{Fisher:1978ww} can be used to characterize the phase transition and probe finite-size scaling \cite{Ganshin:2006gq,*Zandi:2004cv}, while the thermal Casimir effect \cite{Sushkov:2011tl} has been used to probe phase transitions \cite{Ziherl:1998ep}.

In this letter, we consider how the Casimir force changes as we tune a two-dimensional spin-orbit coupled material through a Lifshitz transition. 
A Lifshitz transition occurs when a material's Fermi surface undergoes a topological change -- such as the emergence or collapse of an electron or hole pocket \cite{Lifshitz:1960ux,*Abrikosov:sl2RA27H}. 
Various models are suspected to undergo some type of Lifshitz transition \cite{Chen:2012jo,*Okamoto:2010dw,*Hackl:2011kf} including the cuprates \cite{Norman:2010dj}, and experimental evidence of a Lifshitz transition has been recently observed in iron arsenic superconductors \cite{Liu:2010hp}.
We will first define our model and show how it undergoes such a transition. 
Introducing the expression for the Casimir energy, we then find the current-current correlator in linear response theory after minimally coupling our Hamiltonian to a vector potential.
Using this expression, we numerically integrate to obtain the Casimir force as we tune our original Hamiltonian through a Lifshitz transition.
We end with some discussion of this feature.

\begin{figure}
\centering
\includegraphics[width=\columnwidth]{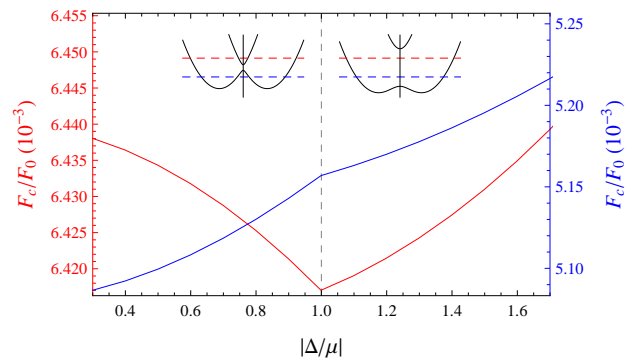} 
\caption{ \label{fig:MetalResults} (Color online) The Casimir force $F_c$ normalized by the ideal conductor value $F_0 = -\hbar c \pi^2/240 a^4$ between one semiconductor plate and one metallic plate separated by a = \unit[20]{nm} as a function of the Zeeman splitting normalized by the Fermi energy. 
The red plot (left axis) corresponds to the Fermi energy crossing the upper band for $|\Delta| < |\mu|$, and the blue plot (right axis) corresponds to the Fermi energy only ever crossing the lower band. The insets show the band structure above and below the transition along with the two fixed values of the Fermi energy.}
\end{figure}

Others have considered the Casimir effect with two-dimensional plates \cite{Parashar:2012it}, however our particular model requires a more material-centered approach (see the supplement \cite{Supplement}). 
We consider the Casimir force at zero temperature between two parallel plates where at least one is modeled as a two-band spin-orbit coupled material (sufficiently thin to be considered quasi-two dimensional) with a fixed chemical potential and tunable Zeeman splitting due to an external magnetic field. 
(When considering only one spin-orbit coupled plate, the other is a metallic plate, modeled as a clean free electron gas.)
The Zeeman field tunes a gap in this two-band material and causes one of the Fermi surfaces to form or collapse. 
This is the simplest realistic model exhibiting a Lifshitz transition and gives at least a qualitative idea of what would happen with a strongly spin-orbit coupled semiconducting material like indium antimonide. 
At these transition points, the Casimir force between the two plates experiences a kink, as seen in Fig.~\ref{fig:MetalResults}.

This could be experimentally measured with the usual plate and sphere geometry as seen in Fig.~\ref{fig:CasimirExperiment}.
The plate would be a thin layer of InSb while the sphere would be the usual Au-coated sphere.
While we consider the parallel plate scenario, our calculations can be generalized to the sphere-plate geometry by using the proximity force approximation \cite{Bordag:2001tt}.

We consider the single-particle effective Hamiltonian for the conduction bands of the semiconductor,
\begin{equation} \label{eq:hamiltonian}
\hat{H} = \frac{k^2}{2m^\ast} - \mu + \beta(\hat{\sigma}_x k_x - \hat{\sigma}_y k_y) + \Delta \hat{\sigma}_z,
\end{equation}
which has eigenvalues
\begin{equation} \label{eq:bands}
\xi^{\pm}(k) = \frac{k^2}{2m^\ast}-\mu \pm \sqrt{\Delta^2 + \beta^2 k^2},
\end{equation}
where $m^\ast$ and $\mu$ are the conduction band effective mass of the electron and chemical potential. 
The coefficient $\beta$ is the strength of the Dresselhaus spin-orbit coupling, and $\sigma_i$ are the Pauli matrices. 
The factor $\Delta$ is the induced Zeeman splitting, given by $\Delta = \mu_B g^\ast B$, where $\mu_B$ is the Bohr magneton, $g^\ast$ is the material's $g$-factor, and $B$ is an applied magnetic field. 
For all calculations we will assume that this Hamiltonian is a simple model of the the relevant bands of the material indium antimonide, for which $m^\ast = 0.014 m_0$, where $m_0$ is the free electron mass, and $\beta = \gamma \langle k_z^2\rangle \simeq \gamma \left(\frac{\pi}{d}\right)^2$ \cite{Koralek:2009bu}, where $d$ is the thickness of the plate and $\gamma = \unit[760.1]{eV \text{\AA}^3}$ is the intrinsic Dresselhaus parameter for the material. 
We consider InSb plates that are six lattice constants thick, $d=6 \times \unit[0.6479]{nm} = \unit[3.89]{nm}$.
The plates may still be considered effectively 2D as long as the energy needed to excite higher electron modes in the confined direction is much larger than the energy requires to excite the two lowest bands modeled here.
Additionally, since the $g$-factor of InSb is $g^\ast = -51.6$ we can also neglect the orbital coupling of the electrons directly to the external magnetic field as well as the effect of the magnetic field on the metallic plate when it is considered \cite{Winkler:2003dc}. 

This model is a simplification since we neglect virtual excitations in the confined direction. 
The force should be dominated by the two bands considered in Eq.~(\ref{eq:bands}), and we expect that at worst, effects due to the confined direction and crystalline structure of InSb to change the quantitative nature, but not the qualitative features we find.

For $\mu > |\Delta|$ there are two bands crossing the Fermi energy. 
Fixing $\mu$, as $|\Delta|$ is increased the occupation of the upper band decreases until the Fermi surface disappears entirely when $|\mu| = |\Delta|$ -- the electron pocket defined by that Fermi surface disappears. 
Increasing the Zeeman splitting further, the Fermi energy lies within the gap and only the lower band crosses the Fermi level, giving a single Fermi surface. 
This represents the Lifshitz transition in the region of $\mu>|\Delta|$, and is shown with the red dashed line in the insets of Fig.~\ref{fig:MetalResults}. 

If $m^\ast \beta^2 > |\Delta|$ the lower band has a local maximum at $k=0$ and a similar scenario can be considered for $\epsilon_{\mathrm{min}}<\mu < -|\Delta|$, where $\epsilon_{\mathrm{min}}$ is the lowest energy of the lower band. 
In this case, the lower band crosses the Fermi energy for two distinct values of $k$, producing two Fermi surfaces -- the inner one enclosing a hole pocket. 
Again, increasing $|\Delta|$ for fixed $\mu$ leads to a shrinking of the inner Fermi surface until it disappears completely at the point when $|\mu| = |\Delta|$. 
For larger Zeeman splitting, the Fermi energy again lies within the gap and there is a single Fermi surface. 
This is shown with the blue dashed line in the insets of Fig.~\ref{fig:MetalResults}. 
The disappearance of a Fermi surface by changing $\Delta$ in these two scenarios are simple examples of a Zeeman-driven Lifshitz transition. 

We use a microscopic quantum field theoretic method to calculate the Casimir energy at zero temperature in terms of the current-current correlation functions of the two electron systems under consideration and virtual photons in the 3D vacuum between them (see supplement \cite{Supplement}). 
Summing up the diagrams in Eq.~\eqref{eq:schematic-casimir}, the Casimir energy at zero temperature for parallel 2D plates separated by a distance $a$ is given by
\begin{widetext}
\begin{equation} \label{eq:casimirenergy}
\mathcal{E}_c(a) = \frac{1}{8\pi^2} \int_0^\infty dq_{\perp} \,q_\perp \int_{-q_\perp}^{q_\perp} d\omega \, \tr \ln \left[ \hat{\bm{1}} - \hat{\widetilde{\Pi}}_A(q_\perp,i\omega) \hat D(q_\perp,i\omega,a) \hat{\widetilde{\Pi}}_B(q_\perp,i\omega) \hat D(q_\perp,i\omega,a) \right],
\end{equation}
\end{widetext}
where $\hat D$ is the photon propagator and $\hat{\widetilde{\Pi}}_i$ is the current-current correlation function for plate $i$, dressed by interactions with 3D photons. 
We choose the axial gauge with $\phi = 0$, so the relevant components of the photon propagator have the form
\begin{equation*}
\hat D(q_\perp,i\omega,z)= \left( \begin{array}{cc}
\frac{q_\perp}{\omega^2} & 0 \\
0 & \frac{1}{q_\perp} \end{array} \right) e^{-q_\perp |z|}.
\end{equation*}
The dressed current-current correlation function can be expressed in terms of the bare correlation function, $\hat{\Pi}$, as
\begin{equation*}
\hat{\widetilde{\Pi}} = \left[\hat{\bm{1}}-\hat{\Pi}\hat{D}(z=0)\right]^{-1}\hat{\Pi},
\end{equation*}
which accounts for dynamical screening of photons in the random phase approximation (RPA). 
We determine the bare correlation function using the current operator, $j_i(x) = \psi^\dagger(x)\frac{\partial \hat{H}[A]}{\partial A_i(x)}\psi(x)$, where $\hat{H}[A]$ is the Hamiltonian given in Eq.~\eqref{eq:hamiltonian} after minimal coupling. 
The correlation function is then expressed in terms of the current as,
\begin{equation} \label{eq:correlation}
\Pi_{ij}(x,x') = \langle-\delta(x-x')\delta_{ij}\partial_{A_i}j_i(x)+j_i(x)j_j(x') \rangle \Big\vert_{A=0},
\end{equation}
where $\langle \cdots \rangle$ represents averaging over the ground state \cite{Altland:2010ww}.
The first term is the diamagnetic term while the second is the paramagnetic term. 
The inclusion of the diamagnetic term is important since in the clean electron system without spin-orbit coupling, it corresponds to the only contribution to conductivity.
In the case of a weakly correlated system we can use the approximation that the Casimir effect is determined by the local current-current response functions, i.e. we only need to consider the $q=0$ limit of $\hat{\Pi}$ since non-local behavior is screened out. 
Furthermore, coupling of the spin to the magnetic fluctuations of the vacuum field do not need to be considered.
In this limit, the correlation function for the spin-orbit coupled plates has the form
\begin{equation}
\hat{\Pi}(i\omega) = -\alpha \left( \begin{array}{cc}
\Pi_L(i\omega) & \Pi_H(i\omega) \\
-\Pi_H(i\omega) & \Pi_L(i\omega) \end{array} \right),
\end{equation}
where $\alpha$ is the fine structure constant,
\begin{align}
\Pi_H(i\omega) &= 2\Delta \left[\cot^{-1}\left(\frac{\omega}{2\epsilon^+}\right) -\cot^{-1}\left(\frac{\omega}{2\epsilon^-}\right) \right] \\
\Pi_L(i\omega) &= 2\mu\left[\Theta(\mu - |\Delta|)+\Theta(\mu+|\Delta|)\right] \\ & \qquad + \epsilon^+ -\epsilon^- + \frac{\omega^2 -4\Delta^2}{4\Delta^2\omega}\Pi_H(i\omega) \nonumber
\end{align}
and $\epsilon^\pm$ are the positive square roots of
\begin{multline}
(\epsilon^\pm)^2 = \Delta^2 + \\
\max \left\{0,2m^\ast \beta^2(\mu + m^\ast \beta^2)\left[1\pm\sqrt{1-\tfrac{\mu^2-\Delta^2}{(\mu+m^\ast \beta^2)^2}}\right]\right\}.
\end{multline}

\begin{figure}
\centering
\includegraphics[width=\columnwidth]{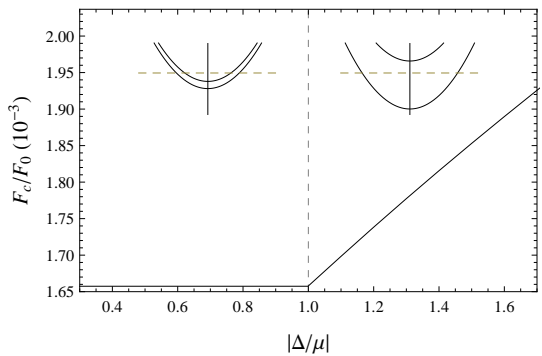} 
\caption{ \label{fig:Beta=0Results} The Casimir force $F_c$ normalized by the ideal conductor value $F_0$ between two metallic plates separated by a = \unit[20]{nm} as a function of the Zeeman splitting normalized by the Fermi energy. The insets show the band structure above and below the transition along with the fixed value of the Fermi energy.}
\end{figure}

We take the derivative of Eq.~\eqref{eq:casimirenergy} with respect to the plate separation, $a$, to obtain an expression for the Casimir force. 
We then integrate numerically for fixed separation $a=\unit[20]{nm}$ and Fermi energy $\mu$, while varying $|\Delta|$, which would correspond to varying the magnetic field in an actual experiment. 
For all numerical results, we will give the Casimir force in our considered system, $F_c$, normalized by the Casimir force between ideal conducting plates, $F_0 = -\hbar c \pi^2/240 a^4$, calculated for the same plate separation.  
For the simple system with no spin-orbit coupling ($\beta=0$), i.e. two metallic plates, we obtain the results in Fig.~\ref{fig:Beta=0Results}. 
In this case, there is only the possibility of a Lifshitz transition associated with removing the electron pocket of the upper band. 
We see that for $|\Delta|<|\mu|$ the Casimir force is constant with varying $|\Delta|$, since the carrier density of the material, which in this case is the only parameter determining the value of $\hat\Pi$, is constant in this region. 
As the upper band is raised above the Fermi level, the closing of the upper band Fermi surface is indicated by a kink in the Casimir force, above which the magnitude of the force increases with $|\Delta|$, consistent with the increase in the carrier density in this region. 
Deviations from this behavior in the case of $\beta\neq 0$ are primarily due to spin-orbit effects. 

\begin{figure}
\centering
\includegraphics[width=\columnwidth]{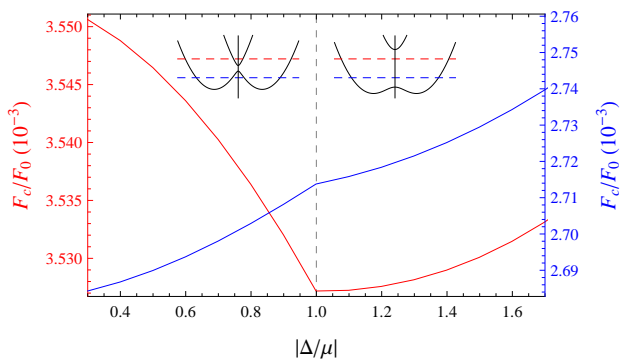} 
\caption{ \label{fig:DresselhausResults} (Color online) The Casimir force $F_c$ normalized by the ideal conductor value $F_0$ between two semiconductor plates separated by a = \unit[20]{nm} as a function of the Zeeman splitting normalized by the Fermi energy. The red plot (left axis) corresponds to the Fermi energy crossing the upper band for $|\Delta| < |\mu|$, and the blue plot (right axis) corresponds to the Fermi energy only ever crossing the lower band. The insets show the band structure above and below the transition along with the two fixed values of the Fermi energy.}
\end{figure}

We find that the Casimir force as a function of Zeeman energy has the same overall features whether the system we consider is one InSb plate and one metallic plate or two InSb plates, shown in Fig.~\ref{fig:MetalResults} and Fig.~\ref{fig:DresselhausResults} respectively. 

For the Fermi energy lying in the upper band, the Casimir force decreases as the Zeeman energy increases until the point when the Fermi surface for the upper band disappears at the transition point, $|\Delta| = |\mu|$, above which the force increases again.
At the Lifshitz transition point, the value of the normalized Casimir force, $F_c/F_0$, is found to be $6.42 \times 10^{-3}$ and $3.53 \times 10^{-3}$ in the semiconductor/metal system and two semiconductor system respectively (numerical results will always be given in this order). 
For $|\Delta|$ just below the transition point, the rate of the Casimir force's change with $|\Delta|$, i.e.\ $\tfrac{d\left(F_c/F_0\right)}{d\Delta}$, is  approximately $-4.5 \times 10^{-3} \unit{eV}^{-1}$ and $-5.0 \times 10^{-3} \unit{eV}^{-1}$ in the two respective cases. 
For $|\Delta|$ just above the transition point, the value of this quantity is approximately $1.9 \times 10^{-3} \unit{eV}^{-1}$ and $ 0.02 \times 10^{-3} \unit{eV}^{-1}$ respectively in the two systems.

We can understand these features intuitively by first examining the carrier density in the spin-orbit coupled material on either side of the transition.
Below the transition, the carrier density remains constant with changing $|\Delta|$ as it does in the $\beta=0$ case, so the decrease in the force is an indication that spin-orbit effects in these two systems, i.e.\ virtual inter-band excitations of electrons near the Zeeman gap, work to weaken the strength of the Casimir force. 
Above the transition, the carrier density increases with increasing $|\Delta|$, which leads to an increase in the strength of the force as in the $\beta=0$ case considered earlier. 
However, we also see that the Casimir force increases much faster just above the transition in Fig.~\ref{fig:MetalResults} than in Fig.~\ref{fig:DresselhausResults}, where the curve is nearly flat.
This suggests further that virtual inter-band spin-orbit interactions, which have a greater effect in the system with two semiconducting plate and are strongest right around the Lifshitz transition point, are responsible for suppressing the force above the transition until the Zeeman gap is large enough to sufficiently dampen the effects. 

For the Fermi energy below the Zeeman gap, as $|\Delta|$ is increased the magnitude of the Casimir force is found to constantly increase. 
There is a noticeable kink in the force as a function of Zeeman field at $|\Delta| = |\mu|$, the point when the inner Fermi surface closes. 
At the transition point, the value of the normalized Casimir force is found to be $5.16 \times 10^{-3}$ between one semiconducting plate and one metallic plate and $2.71 \times 10^{-3}$ between two semiconducting plates. 
For this Fermi level, the carrier density is constantly increasing with increasing $|\Delta|$, so the force is constantly increasing as well. 

As we have shown, tuning through a Lifshitz transition in this material causes a kink in the Casimir force while the microscopics control the nature and severity of the kink. 
We expect similar features to be found in other materials with such transitions -- particularly due to the change in the carrier concentration across such a transition.
As such, the precision Casimir force experiments could be used as a probe of interesting material properties.

\emph{Acknowledgements} -- This work was supported by DOE-BES DESC0001911 (A.A. and V.G.), JQI-PFC (J.W.), and the Simons Foundation. 
We acknowledge useful discussions with Jeremy Munday, Doron Bergmann, and Jing Xia.

\bibliography{references}

\onecolumngrid

\hspace{12pt}

\hrule

\hspace{12pt}
\stepcounter{myequation}

\section{Supplementary material: Non-analytic behavior of the Casimir force across a Lifshitz transition in a spin-orbit coupled material}


\providecommand{\qint}[2]{\int\frac{d^{\,#2} #1}{(2\pi)^{#2}}}
\providecommand{\xint}[2]{\intd^{\,#2} #1}

\section*{Expression for the Casimir Energy}
We wish to derive an expression for the electromagnetic Casimir energy between two parallel two-dimensional plates in terms of photon propagators and quantities that can be derived from the microscopic description of the electrons in each plate. We do this by calculating the free energy of the two plate system interacting with three-dimensional photons, then subtracting off the contribution for each isolated plate and the photon background, leaving only the part that depends on the distance between the plates,
\begin{align*}
\mathcal{E}_c &= F - F_{1} - F_{2} \\
&= -\frac{1}{\beta}\left(\ln \mathcal{Z} -\ln \mathcal{Z}_{1} - \ln \mathcal{Z}_{2} \right) \\
&= -\frac{1}{\beta}\left(\ln \int \mathcal{D}\!A \, e^{-S_{EM}\left[A\right]-S_1\left[A\right]-S_2\left[A\right]} - \ln \int \mathcal{D}\! A' \, e^{-S_{EM}\left[A'\right]-S_1\left[A'\right]} - \ln \int \mathcal{D}\! A'' \, e^{-S_{EM}\left[A''\right]-S_2\left[A''\right]} \right).
\end{align*}
Here, $F$ and $\mathcal{Z}$ are the free energy and partition function of the full system composed of two plates interacting with three-dimensional photons, while $F_i$ and $\mathcal{Z}_i$ are the free energy and partition function of plate $i$ interacting with three-dimensional photons in isolation. The partition functions are calculated as path integrals over the photon field $A$, with the actions for the photon field given by 
\begin{equation*}
S_{EM}\left[A\right] = -\frac{1}{4} \int d z \sum_n \qint{q}{2} \, F_{\mu \nu}\left(\vec q, \omega_n, z \right)F^{\mu \nu}\left(\vec q, \omega_n, z \right), \label{eq:EMaction-arxiv}
\end{equation*}
and the action for the electrons in plate $i$ given by
\begin{equation*}
S_i\left[A\right] = -\frac{1}{2} \int d z\, \delta\left(z-z_i\right)  \sum_{n} \qint{q}{2} \, A^\mu \left(\vec q, \omega_n, z \right)\Pi_{\mu \nu, i}\left(\vec q, \omega_n \right) A^{\nu}\left(-\vec q, -\omega_n, z \right). \label{eq:plateaction-arxiv}
\end{equation*}
Throughout we have also set $\hbar=c=1$. The two plates are located at $z=z_1$ and $z=z_2$ with $z_2-z_1 = a$, the distance between the plates. This expression for $S_i$ is obtained by considering the full action of the electrons in plate $i$ minimally coupled to three-dimensional photons, then integrating out the electronic degrees of freedom and keeping terms only second order in $A$. This procedure gives the finite temperature current-current correlation function from linear response theory. In principle, higher order terms with two external photon lines, but also including internal photon lines, could also be included, but they would be higher order in $\alpha = \tfrac{e^2}{4\pi}\approx \frac1{137}$ and therefore provide only small corrections. 

For each plate, we define the current operator in the usual way, $j_\mu = \frac{\delta}{\delta A^\mu}S [\bar\psi,\psi,A] = \bar\psi \frac{\partial \hat{H}[A]}{\partial A^\mu}\psi$, where $\hat{H}[A]$ is the single particle electronic Hamiltonian for the chosen plate. We can then write an expression for $\hat\Pi$ in terms of this object.
\begin{equation} \label{eq:correlation-arxiv}
\Pi_{\mu\nu}(x,x') = \left\langle -\delta(x-x')\delta_{\mu\nu}\frac{\partial j_{\mu}(x)}{\partial A_{\mu}(x)} + j_\mu(x) j_\nu(x')\right\rangle \Biggr\vert_{A=0}
\end{equation}
where $\langle\cdots\rangle$ here denotes integration over the electron fields \cite{Altland:2010ww-arxiv}.

Introducing the notation
\begin{equation*}
\int \mathcal{D}\!A \, (\cdots) e^{-S_{EM}\left[A\right]} = \langle \, \cdots \rangle_A,
\end{equation*}
we now have that the Casimir energy can be written as,
\begin{align*}
\mathcal{E}_c &= -\frac{1}{\beta}\left(\ln \left\langle e^{-S_1[A]-S_2[A]}\right\rangle_A - \ln \left\langle e^{-S_1[A']}\right\rangle_{A'} - \ln \left\langle e^{-S_2[A'']}\right\rangle_{A''}\right) \\
&= -\frac{1}{\beta}\left(\ln \left\langle e^{-S_1[A]-S_2[A]}\right\rangle_A - \ln \left\langle e^{-S_1[A']-S_2[A'']}\right\rangle_{A',A''}\right) 
\end{align*}
A careful expansion of all the terms, after many cancellations and renaming $A'$ and $A''$ to $A$ where the distinction is unnecessary, gives
\begin{multline*}
\mathcal{E}_c = -\frac{1}{\beta} \left\{\left(\langle S_1S_2\rangle_A - \langle S_1\rangle_A\langle S_2\rangle_A\right) - \left(\frac{1}{2}\langle S_1^2S_2\rangle_A - \frac{1}{2} \langle S_1^2\rangle_A \langle S_2\rangle_A -\langle S_1\rangle_A\langle S_1S_2\rangle_A - \langle S_1\rangle_A^2\langle S_2\rangle_A \right.\right.\\ 
\left.\left.+ \frac{1}{2}\langle S_1S_2^2\rangle_A - \frac{1}{2} \langle S_1\rangle_A \langle S_2^2\rangle_A -\langle S_2\rangle_A\langle S_1S_2\rangle_A - \langle S_1\rangle_A\langle S_2\rangle_A^2 \right) +\dots \right\}.
\end{multline*}
All remaining terms involve both plates in the form of $S_1$ and $S_2$. Terms that are the product of averages, e.g. $\langle S_1\rangle\langle S_2\rangle$, represent disconnected diagrams, with each part assocciated with only a single plate, and will cancel the corresponding disconnected parts of terms involving the average of products, e.g. $\langle S_1S_2\rangle$, so that only connected diagrams involving both plates remain. Diagrammatically, we are left with two types of connected diagrams: 
\begin{figure}[h] \label{fig:diagrams-arxiv}
\centering
\subfloat[]{
\includegraphics[scale=0.4]{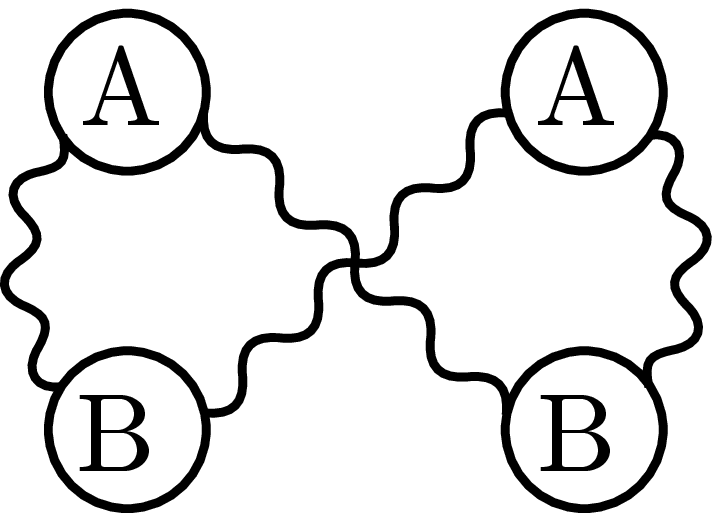}
\label{fig:crossdiagram-arxiv}
}
\qquad \qquad
\subfloat[]{
\includegraphics[scale=0.4]{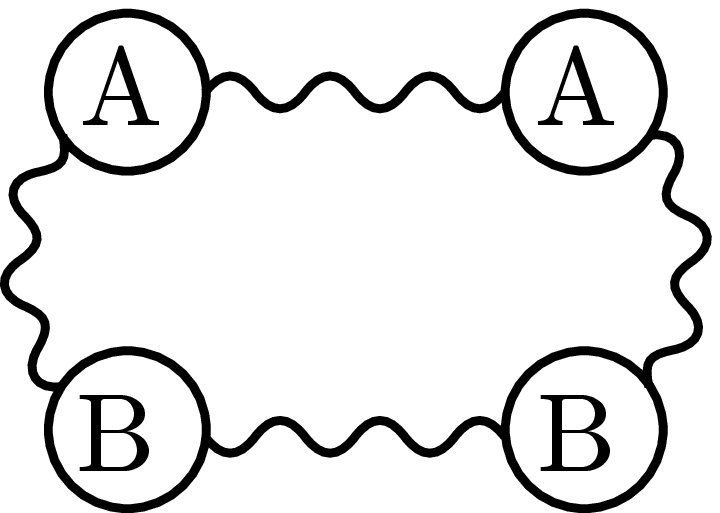}
\label{fig:loopdiagram-arxiv}
}
\caption{Examples of the two types of diagrams we are left with. In the first type, shown in (a), all photon lines connect one plate (in the form of $\hat{\Pi}_1$) to the other plate (in the form of $\hat{\Pi}_2$). In the second type, shown in (b), there is at least a single photon line connecting a plate to itself.}
\label{fig:RashbaCasimir-arxiv}
\end{figure} 

The type of diagram demonstrated in Fig.~\ref{fig:loopdiagram-arxiv} dresses the interaction of the photons with the plates. Keeping track of the coefficient of every such term that leads to these diagrams in the expansion above and calculating the symmetry factors of such diagrams, we find that we may define the photon-dressed current-current correlation function as
\begin{equation} \label{eq:dressedPi-arxiv}
\hat{\widetilde{\Pi}} = \hat\Pi + \hat\Pi \hat D(z=0) \hat\Pi + \hat\Pi \hat D(z=0) \hat\Pi \hat D(z=0) \hat\Pi + \dots = \left(\hat{\bm{1}}-\hat\Pi\hat D(z=0)\right)^{-1}\hat\Pi
\end{equation}
and we are left with only the first type of diagrams, but with $\hat\Pi$ replaced with $\hat{\widetilde\Pi}$. This is equivalent to considering the random phase approximation (RPA) for the total current-current correlation.  Using this definition and calculating the appropriate coefficient for each diagram, we can write a single expression that incorporates all connected diagrams. 
\begin{align*}
\mathcal{E}_c(a) &= -\frac{1}{\beta} \sum_n \qint{q}{2} \tr \left[\sum_{k=1}^\infty \frac{1}{2k}\left( \hat{\widetilde{\Pi}}_1(\vec q,\omega_n) \hat D(\vec q,\omega_n,a) \hat{\widetilde{\Pi}}_2(\vec q,\omega_n) \hat D(\vec q,\omega_n,a)\right)^k\right] \\
&= \frac{1}{2\beta}\sum_n \qint{q}{2} \tr \ln \left[\hat{\bm{1}} - \hat{\widetilde{\Pi}}_1(\vec q,\omega_n) \hat D(\vec q,\omega_n,a) \hat{\widetilde{\Pi}}_2(\vec q,\omega_n) \hat D(\vec q,\omega_n,a)\right] \\
& \overset{T\to 0}{=} \frac{1}{2} \int_{-\infty}^{\infty} \frac{d \omega}{2\pi} \qint{q}{2} \tr \ln \left[\hat{\bm{1}} - \hat{\widetilde{\Pi}}_1(\vec q,i\omega) \hat D(\vec q,i\omega,a) \hat{\widetilde{\Pi}}_2(\vec q,i\omega) \hat D(\vec q,i\omega,a) \right]
\end{align*}
In the limit $T\to 0$, we have not performed the analytic continuation $i\omega_n \to \omega + i0^+$, we simply make the discrete Matsubara sum into a continuous integral over imaginary frequency. Assuming that the integrand depends only on the magnitude of $\vec q$ and not its direction, we may perform the angular integration. We now perform a change of variables. Define $q_\perp = \sqrt{\omega^2 + q^2}$. This would have the form of the perpendicular component of the momentum for on-shell photons if we completed the analytic continuation, but here it is simply a formal change of variables that makes the expression calculationally simpler. With this change of variables and a reordering of integrals we arrive at a final expression for the Casimir energy: 
\begin{equation} \label{eq:casimir-arxiv}
\mathcal{E}_c \left( a \right) = \frac{1}{8 \pi^2} \int_0^\infty d q_{\perp} \,q_\perp \int_{-q_\perp}^{q_\perp} d \omega \, \tr \ln \left[ \hat{\bm{1}} - \hat{\widetilde{\Pi}}_1( q_\perp,i\omega) \hat D(q_\perp,i\omega,a) \hat{\widetilde{\Pi}}_2(q_\perp,i\omega) \hat D(q_\perp,i\omega,a) \right]
\end{equation}
We choose the axial gauge with $\phi = 0$, so the relevant components of the photon propagator have the form
\begin{equation*}
\hat D(q_\perp,i\omega,z)= \left( \begin{array}{cc}
\frac{q_\perp}{\omega^2} & 0 \\
0 & \frac{1}{q_\perp} \end{array} \right) e^{-q_\perp |z|}.
\end{equation*}
Since one may define the frequency dependent conductivity tensor as $\hat\sigma(\omega) = -\hat\Pi / \omega_n\vert_{i\omega_n\to\omega+i0^+}$, we may also define the photon-dressed conductivity tensor (evaluated at imaginary argument) as 
\begin{equation*}
\hat{\widetilde\sigma}(i\omega)=-\frac{\hat{\widetilde\Pi}}{\omega} = -\left(\hat{\bm{1}}-\hat\Pi \hat D\right)^{-1}\frac{\hat\Pi}{\omega} = \left(\hat{\bm{1}} + \omega \hat\sigma \hat D\right)^{-1} \hat\sigma
\end{equation*}
Using this definition, we can write an alternative form of the expression for the Casimir energy:
\begin{equation} \label{eq:conductivitycasimir-arxiv}
\mathcal{E}_c \left( a \right) = \frac{1}{8 \pi^2} \int_0^\infty d q_{\perp} \,q_\perp \int_{-q_\perp}^{q_\perp} d \omega \, \tr \ln \left[ \hat{\bm{1}} - \omega^2\,\hat{\widetilde{\sigma}}_1(q_\perp,i\omega) \hat D(q_\perp,i\omega,a) \hat{\widetilde{\sigma}}_2(q_\perp,i\omega) \hat D(q_\perp,i\omega,a) \right]
\end{equation}
This form is sometimes favorable since it is explicitly in terms of the longitudinal ($\sigma_{xx}$ and $\sigma_{yy}$) and Hall ($\sigma_{xy}$) conductivities of the plates.

We can check the validity of this equation by applying it to an analytically tractable system seeing if we arrive at the correct result. To do so, we will need a systematic way to analytically work with Eq.~(\ref{eq:casimir-arxiv}).
First, we rewrite the trace-log using the identity $\tr\ln \hat M = \ln\det\hat M$ for any non-singular matrix $\hat M$, assuming that $\hat{\bm{1}}-\hat{\widetilde\Pi}_1 \hat D \hat{\widetilde\Pi}_2 \hat D$ is non-singular for all $q_\perp$ and $\omega$. Then we use the fact that the determinant of a matrix is equal to the product of its eigenvalues. Defining the components of $\hat{\widetilde\Pi}_1 \hat D \hat{\widetilde\Pi}_2 \hat D$ as
\begin{equation*}
\hat{\widetilde\Pi}_1 \hat D \hat{\widetilde\Pi}_2 \hat D = \left(\begin{array}{cc} m_{11} & m_{12} \\
m_{21} & m_{22} \end{array}\right)e^{-2q_\perp a}
\end{equation*}
we can write the eigenvalues of $\hat{\bm{1}}-\hat{\widetilde\Pi}_1 \hat D \hat{\widetilde\Pi}_2 \hat D$ as:
\begin{align*}
\lambda_\pm &= 1-\frac{1}{2}\left(\tr\left[\hat{\widetilde\Pi}_1 \hat D \hat{\widetilde\Pi}_2 \hat D\right] \pm \sqrt{\tr\left[\hat{\widetilde\Pi}_1 \hat D \hat{\widetilde\Pi}_2 \hat D\right]^2 -4\det\left[\hat{\widetilde\Pi}_1 \hat D \hat{\widetilde\Pi}_2 \hat D\right]} \right) \\
&= 1-\left(\frac{m_{11}+m_{22}}{2}\pm \sqrt{\left(\frac{m_{11}-m_{22}}{2}\right)^2 + m_{12}m_{21}}\right) e^{-2q_\perp a} \\ 
&= 1-\kappa_\pm(q_\perp,i\omega) e^{-2q_\perp a}
\end{align*}
So we can rewrite Eq.~(\ref{eq:casimir-arxiv}) in the form:
\begin{align}
\mathcal{E}_c \left( a \right) &= \frac{1}{8 \pi^2} \int_0^\infty d q_{\perp} \,q_\perp \int_{-q_\perp}^{q_\perp} d \omega \, \tr \ln \left[ \hat{\bm{1}} - \hat{\widetilde{\Pi}}_1(q_\perp,i\omega) \hat D(q_\perp,i\omega,a) \hat{\widetilde{\Pi}}_2(q_\perp,i\omega) \hat D(q_\perp,i\omega,a) \right] \nonumber \\
&= \frac{1}{8 \pi^2} \int_0^\infty d q_{\perp} \,q_\perp \int_{-q_\perp}^{q_\perp} d \omega \, \left[\ln\left(1-\kappa_+e^{-2q_\perp a}\right) + \ln\left(1-\kappa_-e^{-2q_\perp a}\right)\right] \label{eq:lifshitz-arxiv}\\
&\underset{|\kappa_\pm|<1}{=} -\frac{1}{8 \pi^2} \int_0^\infty d q_{\perp} \,q_\perp \int_{-q_\perp}^{q_\perp} d \omega \, \sum_{n=1}^\infty \frac{\kappa_+(q_\perp,i\omega)^n + \kappa_-(q_\perp,i\omega)^n}{n}e^{-2nq_\perp a} \label{eq:fullexpansion-arxiv}
\end{align}

The assertion that $|\kappa_\pm|<1$ is justified; we observe that Eq.~(\ref{eq:lifshitz-arxiv}) has the same form as the Lifshitz formula for the Casimir energy, and what we are calling $\kappa_\pm$ here are the product of the reflection coefficients for the two plates for the two polarizations of the electromagnetic field there. Since reflection coefficients have to be $\leq 1$, with the equality being for perfect mirrors, we see that $|\kappa_\pm|<1$ is true in general for any non-ideal physical material.

We will now consider the simple case of the Casimir force between two clean two-dimensional free electron gases. The expression for the current-current correlation function of such plates is
\begin{equation*}
  \hat\Pi(i \omega) = \left(\begin{array}{cc}
    - \omega_p & 0 \\ 0 & -\omega_p
  \end{array}\right) , \qquad \omega_p = 2 \alpha \epsilon_F 
\end{equation*}
and we dress it using Eq.~(\ref{eq:dressedPi-arxiv}). Since everything is diagonal, this is easily computed and we obtain
\begin{equation*}
  \hat{\widetilde\Pi}(i\omega)  =  \left(\begin{array}{cc}
      -\frac{\omega_p}{1 + \frac{\omega_p q_\perp}{\omega^2}} & 0 \\ 0 & -\frac{\omega_p}{1 + \frac{\omega_p}{q_\perp}}
    \end{array}\right)
\end{equation*}
We can use Eq.~(\ref{eq:fullexpansion-arxiv}), calculating $\kappa_\pm$ with this $\hat{\widetilde\Pi}$. We obtain,
\begin{align*}
  \mathcal{E}_c(a) = -\frac1{8 \pi^2} \int_0^\infty d q_\perp \, q_\perp \int_{-q_\perp}^{q_\perp} d \omega \, \sum_{n=1}^\infty \left\{\frac1n \left(\frac{\omega_p q_\perp/\omega^2}{1 + \frac{\omega_p q_\perp}{\omega^2}} \right)^{2n}e^{- 2 n q_\perp a} + \frac1n \left(\frac{\omega_p/q_\perp}{1 + \frac{\omega_p}{q_\perp}}\right)^{2n}e^{-2nq_\perp a} \right\},
\end{align*}
Now, we can exchange sum and integrals, and let $x = 2 n q_\perp a$ and $y= 2 n \omega a$ to obtain
\begin{align*}
  \mathcal{E}_c(a) = -\frac1{8 \pi^2}  \sum_{n=1}^\infty \frac1{(2 n a)^3} \int_0^\infty d x \, x \int_{-x}^{x} d y \,\left\{\frac1n \left(\frac{2 n a \omega_p x}{y^2 + 2 n a \omega_p x} \right)^{2n}e^{- x} + \frac1n \left(\frac{2 n a \omega_p}{x + 2 n d \omega_p}\right)^{2n}e^{-x} \right\}.
\end{align*}
We can actually perform the $y$ integral to obtain
\begin{align*}
  \int_{-x}^{x} d y \left(\frac{2 n a \omega_p x}{y^2 + 2 n a \omega_p x} \right)^{2n} & = 2 x \, \tensor[_2]{F}{_1}[\tfrac12,2n;\tfrac32; -\tfrac{x}{2n a \omega_p}] \\ & = 2  \sum_{k = 0}^\infty \frac1{2k+1} \frac{(2n+k-1)!}{(2n-1)!} \frac{(-1)^k}{k!} \frac{x^{k+1}}{(2 n a \omega_p)^k} \\
 2x \left(\frac{2 n a \omega_p}{x + 2 n a \omega_p}\right)^{2n} & = 2 \sum_{k=0}^\infty \frac{(2n+k-1)!}{(2n-1)!} \frac{(-1)^k}{k!} \frac{x^{k+1}}{(2na \omega_p)^k}.
\end{align*}
Now, all integrals can be done and we have
\begin{equation}
\label{eq:asymptoticplasma-arxiv}
  \mathcal{E}_c(a) = -\frac1{16 \pi^2 d^3}    \sum_{k = 0}^\infty \frac{(k+2)(k+1)^2}{2k+1}\left[ \sum_{n=1}^\infty \frac1{n^4}\frac1{(2n)^k} \frac{(2n+k-1)!}{(2n-1)!}\right] (-1)^k \frac{1}{(a \omega_p)^k}.
\end{equation}
This expression is already in an asymptotic form, so we just need to calculate the first couple of terms. The first two terms have exact summed forms:
\begin{equation}
\label{eq:plasmaresult-arxiv}
\mathcal{E}_c(a) = -\frac{\pi^2}{720 a^3} \left[ 1 - 2 \frac1{a \omega_p} + \cdots \right]
\end{equation}
We see here that we find the perfect conductor result for the Casimir energy originally calculated by Casimir \cite{Casimir:1948dh-arxiv}, but with corrections since a two-dimensional electron gas is not synonymous with a perfect conductor. 

\newpage
\section*{CURRENT-CURRENT CORRELATION FUNCTION IN A SPIN-ORBIT COUPLED MATERIAL}
The central object needed for an explicit calculation of the the Casimir energy using this formalism is the current-current correlation function, which encodes the interaction of the photon field with the plates to order $e^2$. With an explicit form of this function for each plate, it becomes simple to find the Casimir energy as a function of plate separation by numerically integrating Eq.~(\ref{eq:casimir-arxiv}). Due to screening effects, we only need to consider the local, i.e.\ $q=0$, limit of this correlation function.

We start from Eq.~(\ref{eq:correlation-arxiv}), with the current defined as $j_\mu = \frac{\delta}{\delta A^\mu}S [\bar\psi,\psi,A] = \bar\psi \frac{\partial \hat{H}[A]}{\partial A^\mu}\psi$. Since the gauge we consider in deriving the Casimir effect has $\phi=0$, we will work in that gauge here. This is reflected in the change from $\mu, \nu$ to $i, j$. The diamagnetic term is then
\begin{equation} \label{eq:diamagnetic-arxiv}
\left\langle -\delta(x-x')\delta_{ij}\frac{\partial j_{i}(x)}{\partial A_{i}(x)}\right\rangle\Biggr\vert_{A=0} = \left\langle-\delta(x-x')\delta_{ij}\bar{\psi}(x)\frac{\partial^2 \hat{H}[A]}{\partial A^{i \,2}}\psi(x) \right\rangle\Biggr\vert_{A=0} = -\delta(x-x')\delta_{ij} \tr\left[\hat{G}(x,x) \frac{\partial^2 \hat{H}[A]}{\partial A^{i \,2}}\Biggr\vert_{A=0} \right]
\end{equation}
where trace is over spin indices. Similarly, paramagnetic term is
\begin{equation*}
\left\langle j_i(x) j_j(x')\right\rangle \Big\vert_{A=0} = \left\langle \bar{\psi}(x)\frac{\partial\hat{H}[A]}{\partial A^i}\psi(x) \bar{\psi}(x')\frac{\partial\hat{H}[A]}{\partial A^j}\psi(x') \right\rangle\Biggr\vert_{A=0}
\end{equation*}
Assuming that $\left\langle j_i \right\rangle=0$, i.e., there is no current in equilibrium, then after contracting the electron fields and defining the quantity $\hat{J}_i = \frac{\partial\hat{H}[A]}{\partial A^i}\Big\vert_{A=0}$, we have 
\begin{equation}\label{eq:paramagnetic-arxiv}
\left\langle j_i(x) j_j(x')\right\rangle \Big\vert_{A=0} = -\tr\left[\hat{G}(x,x')\hat{J}_i(x)\hat{G}(x',x)\hat{J}_j(x')\right]
\end{equation}
As before, the trace is over spin indices. Combining Eq.~(\ref{eq:diamagnetic-arxiv}) and Eq.~(\ref{eq:paramagnetic-arxiv}), we arrive at the expression
\begin{equation}
\Pi_{ij}(x,x') = -\delta(x-x')\delta_{ij} \tr\left[\hat{G}(x,x) \frac{\partial^2 \hat{H}[A]}{\partial A^{i \,2}}\Biggr\vert_{A=0} \right] -\tr\left[\hat{G}(x,x')\hat{J}_i(x)\hat{G}(x',x)\hat{J}_j(x')\right]
\end{equation}
We Fourier transform this, giving
\begin{equation}
\Pi_{ij}(\vec q,\omega_n) = -T \sum_m \qint{k}{2} \left\{\delta_{ij}\tr\left[\hat{G}(\vec k,\omega_m) \tfrac{\partial^2 \hat{H}[A]}{\partial A^{i \,2}}\Biggr\vert_{A=0}\right] + \tr\left[\hat{G}(\vec k,\omega_m)\hat{J}_i(\vec k,\vec q)\hat{G}(\vec k+\vec q,\omega_{m+n})\hat{J}_j(\vec k +\vec q,-\vec q)\right] \right\}
\end{equation}
We simplify this using a particular representation of the electron Green's function. 
\begin{equation}
\hat{G}(\vec k,\omega_m) = \sum_\mu \frac{\hat{P}^\mu(\vec k)}{i\omega_m -\xi^\mu_{\vec k}}
\end{equation}
where $\mu$ is now the band index, $\hat P^\mu$ are the projection operators onto the bands and $\xi_{\vec p}^\mu$ are the eigenenergies measured from the chemical potential. For the two-band case of spin-orbit materials, $\mu=\pm$. Inserting this form of the Green's function, we obtain,
\begin{multline*}
\Pi_{ij}(\vec q,\omega_n)=-T\sum_m\qint{k}{2}\left\{ \delta_{ij} \tfrac{\partial^2 \hat{H}[A]}{\partial A^{i \,2}}\Biggr\vert_{A=0} \sum_{\mu} \frac{1}{i\omega_m -\xi_{\vec k}^\mu} + \sum_{\mu,\nu}\frac{1}{i\omega_m-\xi_{\vec k}^\mu}\frac{1}{i\omega_{m+n}-\xi_{\vec k+\vec q}^\nu}\times\right.\\ 
\left.\times\tr\left[\hat P^\mu(\vec k)\hat{J}_i(\vec k,\vec q)\hat P^\nu(\vec k+\vec q)\hat{J}_j(\vec k+\vec q,-\vec q) \right] \right\}
\end{multline*}
Since in general neither the projection operators nor the velocity operators depend on frequency, we can perform the sum over Matsubara frequencies $\omega_m$. 
\begin{equation*}
\Pi_{ij}(\vec q,i\omega) = -\qint{k}{2}\left\{\sum_\mu \delta_{ij} \tfrac{\partial^2 \hat{H}[A]}{\partial A^{i \,2}}\Biggr\vert_{A=0} n_F(\xi_{\vec k}^\mu) + \sum_{\mu,\nu}\frac{n_F(\xi_{\vec k}^\mu)-n_F(\xi_{\vec k+\vec q}^\nu)}{i\omega -\omega_{\vec k \vec q}^{\mu\nu}}\mathcal{F}_{ij}^{\mu\nu}(\vec k,\vec q) \right\}
\end{equation*}
Here we have taken $T\rightarrow 0$, so $\omega_m \rightarrow \omega$, $n_F(x)$ is the Fermi function (which is $\theta(-x)$ at $T=0$), and we have introduced the notation,
\begin{gather*}
\mathcal{F}_{ij}^{\mu\nu}(\vec k,\vec q)=\tr\left[\hat P^\mu(\vec k)\hat{J}_i(\vec k,\vec q)\hat P^\nu(\vec k+\vec q)\hat{J}_j(\vec k+\vec q,-\vec q) \right] \\
\omega_{\vec k \vec q}^{\mu \nu} = \xi_{\vec k+\vec q}^\nu - \xi_{\vec k}^\mu
\end{gather*}
The last step is to take the $\vec q\rightarrow 0$ limit as described before. With this, we arrive as far as we can go without calculating anything specific to the system in question:
\begin{equation}\label{eq:correlationq=0-arxiv}
\Pi_{ij}(i\omega)=\Pi_{ij}(\vec q\rightarrow 0,i\omega) = -\qint{k}{2}\left\{\sum_\mu \delta_{ij} \tfrac{\partial^2 \hat{H}[A]}{\partial A^{i \,2}}\Biggr\vert_{A=0} n_F(\xi_{\vec p}^\mu) + \sum_{\mu,\nu}\frac{n_F(\xi_{\vec k}^\mu)-n_F(\xi_{\vec k}^\nu)}{i\omega -\omega_{\vec k \vec 0}^{\mu\nu}}\mathcal{F}_{ij}^{\mu\nu}(\vec k,\vec 0) \right\}
\end{equation}

For a material with Dresselhaus spin-orbit coupling and a Zeeman field, the Hamiltonian, eigenvectors and eigenvalues are
\begin{equation}
\begin{gathered}
\hat{H} = \frac{k^2}{2m^\ast}\hat{\bm{1}} - \mu + \beta\left(\hat{\sigma}_x k_x - \hat{\sigma}_y k_y\right)+\Delta \hat{\sigma}_z = \left(\begin{array}{cc} \tfrac{k^2}{2m^\ast} + \Delta & \beta k e^{i\theta} \\ \beta k e^{-i\theta} & \tfrac{k^2}{2m^\ast} -\Delta \end{array}\right) \\
\psi^+(\vec k)= \frac{1}{\sqrt{1+\gamma^2}}\left(\begin{array}{c} 1 \\ \gamma \,e^{-i\theta}\end{array}\right) \equiv \left(\begin{array}{c} \cos\phi \\ \sin\phi \,e^{-i\theta}\end{array}\right), \qquad \psi^-(\vec k)= \frac{1}{\sqrt{1+|\gamma |^2}}\left(\begin{array}{c} -\gamma \,e^{i\theta} \\ 1 \end{array}\right) \equiv \left(\begin{array}{c} -\sin\phi \,e^{i\theta} \\ \cos\phi \end{array}\right)\\
\xi_{\vec k}^\pm = \frac{k^2}{2m^\ast} -\mu \pm \sqrt{\Delta^2 + \beta^2 k^2} \equiv \frac{k^2}{2m^\ast} -\mu \pm \lambda_k
\end{gathered}
\end{equation}
where we have defined $k_x \pm i k_y = k\,e^{\pm i\theta}$, $\gamma = \tfrac{\beta \, k}{\Delta + \lambda_k}$, $\cos\phi = \tfrac{1}{\sqrt{1+\gamma^2}}$, and $\sin\phi = \tfrac{\gamma}{\sqrt{1+\gamma^2}}$. From the eigenvectors we can define the projectors onto the bands in the typical way, $\hat{P}^\pm = \psi^\pm \otimes (\psi^\pm)^\dagger$, and we find
\begin{equation}
\hat{P}^+(\vec k) = \left(\begin{array}{cc}\cos^2\phi & \cos\phi\sin\phi \,e^{i\theta} \\
\cos\phi \sin\phi \,e^{-i\theta} & \sin^2\phi \end{array}\right), \qquad \hat{P}^-(\vec k) = \left(\begin{array}{cc}\sin^2\phi & -\cos\phi\sin\phi \,e^{i\theta} \\
-\cos\phi \sin\phi \,e^{-i\theta} & \cos^2\phi \end{array}\right).
\end{equation}
Additionally, we can minimally couple the Hamiltonian to the photon field $A$ and perform the needed derivatives with respect to it. After differentiating and Fourier transforming so that the photon fields carry momentum $\vec q$, we obtain
\begin{gather*}
\frac{\partial^2 \hat{H}[A]}{\partial A^{i \,2}}\Biggr\vert_{A=0} = \frac{e^2}{m^\ast} \\
\hat{J}_i(\vec k,\vec q) = -\frac{e}{2m^\ast}(2\vec k+\vec q)_i\bm{1} - e\beta\hat{\tilde{\sigma}}_i, \qquad \text{where }\hat{\tilde{\sigma}}_x = \hat{\sigma}_x,\quad \hat{\tilde{\sigma}}_y = -\hat{\sigma}_y.
\end{gather*}
Inserting these into Eq.~(\ref{eq:correlationq=0-arxiv}), it is then a straightforward calculation to show that the current-current correlation function is
\begin{equation}
\hat{\Pi}(i\omega) = -\alpha \left( \begin{array}{cc}
\Pi_L(i\omega) & \Pi_H(i\omega) \\
-\Pi_H(i\omega) & \Pi_L(i\omega) \end{array} \right),
\end{equation}
where $\alpha$ is the fine structure constant,
\begin{gather}
\Pi_H(i\omega) = 2\Delta \left[\cot^{-1}\left(\frac{\omega}{2\epsilon^+}\right) -\cot^{-1}\left(\frac{\omega}{2\epsilon^-}\right) \right] \\
\Pi_L(i\omega) = 2\mu\left[\Theta(\mu - |\Delta|)+\Theta(\mu+|\Delta|)\right] + \epsilon^+ -\epsilon^- + \frac{\omega^2 -4\Delta^2}{4\Delta^2\omega}\Pi_H(i\omega),
\end{gather}
and $\epsilon^\pm$ are the positive square roots of
\begin{equation}
(\epsilon^\pm)^2 = \Delta^2 + \max \left\{0,2m^\ast \beta^2(\mu + m^\ast \beta^2)\left[1\pm\sqrt{1-\tfrac{\mu^2-\Delta^2}{(\mu+m^\ast \beta^2)^2}}\right]\right\}.
\end{equation}
These correspond to Eqs. (6-9) in the main text and are used in our numerical analysis. 

\end{document}